# Beyond Acoustics – Capacity Limitations of Linguistic Levels

*Jérémy Giroud[1] & Benjamin Morillon[2]*

[1] *MRC Cognition and Brain Sciences Unit, University of Cambridge, Cambridge, UK*

[2] *Aix Marseille Université, INSERM, INS, Institut de Neurosciences des Systèmes, Marseille, France*

## 1. Speech Processing as Information Transmission through Constrained Channels

Speech is crucial in our daily lives, enabling direct interaction and communication. Understanding speech is a complex process due to its transient and intricately structured nature. Indeed, speech sounds can be abstracted at multiple levels of analysis, speech being a multiplexed signal displaying levels of complexity, organizational principles, and perceptual units of analysis at distinct timescales.

How does the brain build the diverse representational linguistic units at different time scales from the speech signal? This process is even harder given the fleeting nature of speech and human memory limitation. Indeed, speech comprehension faces the "now or never bottleneck" (Christiansen & Chater, 2016). This means that if listeners do not process relevant information in a fast and incremental fashion, they may lose the opportunity to understand it altogether. Therefore, the speed at which information (being acoustic or linguistic in nature) is conveyed in speech – the information rate — is a more relevant dimensional space than the absolute amount of information conveyed (information value) to the brain (Coupé, Oh, Dediu, & Pellegrino, 2019). This is because our neurocognitive resources are limited and can only process a certain amount of information in a given amount of time, leading to temporal bottlenecks (Hasson, Yang, Vallines, Heeger, & Rubin, 2008; Honey et al., 2012; Lerner, Honey, Silbert, & Hasson, 2011; Vagharchakian, Dehaene-Lambertz, Pallier, & Dehaene, 2012). By understanding these bottlenecks and their implications, we can better understand how we process speech.

A potential way of uncovering general principles of speech perception is to describe and determine the temporal constraints that shape its processing at each level of speech analysis. As such, in this chapter, we will show that a meticulous characterization of the various levels of organization found in speech and language, their temporal constraints and their relation to comprehension, can provide valuable and novel insights into an individual's speech processing ability.

Speech processing in the human brain can be conceptualized as a process of information integration through channels with limited capacities. The auditory system continuously receives a complex stream of sound waves that needs to be processed and decoded in real time into meaningful information to understand the messages conveyed by the speaker. This process involves several stages of analysis, from low-level acoustic processing to high-level semantic interpretation (Christiansen & Chater, 2016; Hickok & Poeppel, 2007; Rosen, 1992). One of the key challenges of speech processing is dealing with the limited capacity of our neural resources which results from intrinsic biological constraints. This implies that speech signals or speaking situations that do not conform with these constraints result in poor comprehension.

Building on previous work (Coupé et al., 2019; Ghitza, 2013; Gibson et al., 2019; Pellegrino, Coupé, & Marsico, 2011; Reed & Durlach, 1998), we recently proposed to determine how the limited capacity of our neural resources and the complexity of linguistic features in speech constrain our ability to comprehend spoken language (Giroud et al., 2023). We proposed to rely on a concept inherited from information theory (Shannon, 1948), channel capacity. Within this framework, each level of the speech processing hierarchy can be modeled as a transfer of information through a dedicated channel. Channel capacity is defined as the maximum amount of information that can be transmitted through this communication channel without errors or loss, in bits per second (bits/s). It can be also referred to as a temporal (processing) bottleneck, in which information is processed at a fixed speed (Vagharchakian et al., 2012). Hence, if too much information arrives per unit of time, information transfer is suboptimal or fails. Using such a normative measurement framework allowed for the determination of multilevel linguistic processing constraints limiting speech comprehension. This suggests that speech perception is hierarchical (Dunbar, Pallier, & King, 2022), with sequential bottlenecks, each with its own channel capacity. Hereafter, we will provide an account of diverse experimental works that brought insights about the relevant processing bottlenecks involved in speech comprehension. We will focus on work spanning multiple levels of analysis from acoustic to higher-level linguistic features to determine their respective channel capacity. More precisely, we hereafter characterize the following linguistic features: the speech acoustic time scales; the syllabic time scale; the phonemic time scale; higher-level linguistic time scales such as words, phrases and sentences; lexical information rates and contextual information, as derived from deep neural networks (Figure 1; see also Giroud et al., 2023).

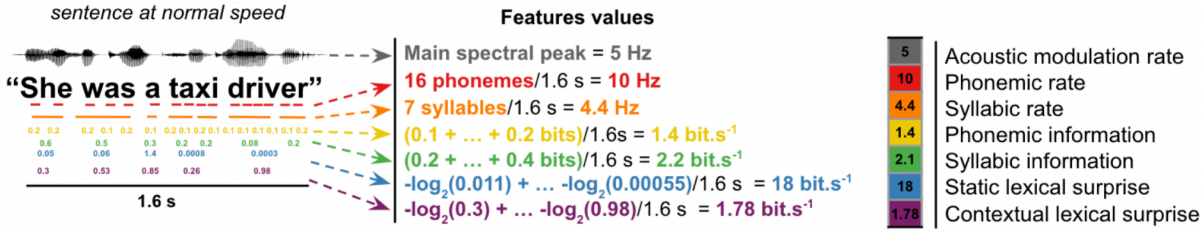

FIGURE 1. SPEECH CHARACTERIZATION AT MULTIPLE LEVELS OF ANALYSIS. RATE (IN HZ) OR INFORMATION RATE (IN BITS/S) OF SEVEN LINGUISTIC FEATURES OF AN EXAMPLE SENTENCE. FEATURES ARE DESCRIBED FROM LOW TO HIGH LINGUISTIC LEVELS: ACOUSTIC TEMPORAL MODULATION RATE (IN HZ), SYLLABIC RATE (IN HZ), PHONEMIC RATE (IN HZ), SYLLABIC INFORMATION RATE (IN BIT/S), PHONEMIC INFORMATION RATE (IN BIT/S), STATIC LEXICAL SURPRISE (I.E., WORD FREQUENCY; IN BIT/S) AND CONTEXTUAL LEXICAL SURPRISE (IN BIT/S).

## 2. THE SPEECH ACOUSTIC TIME SCALES

First and foremost, speech is a complex acoustic signal that involves variations in frequency and intensity over time. These low-level acoustic features can be described in terms of spectro-temporal modulations (Elliott & Theunissen, 2009), and are critical for the intelligibility of speech. On one hand, the spectral (or frequency) dimension is a crucial aspect of the speech signal. It corresponds to the distribution of the energy of the sound signal in the frequency scale (sound spectrum), and makes it possible to define the different formants of the speech units (in particular the vowels) and their transitions (Stevens & Klatt, 1974). On the other hand, the temporal dimension of the speech sounds is highly relevant for comprehension (Albouy, Benjamin, Morillon, & Zatorre, 2020; Shannon, Zeng, Kamath, Wygonski, & Ekelid, 1995; Smith, Delgutte, & Oxenham, 2002). This second dimension indexes the precise organization of the different elements of speech over time.

When producing speech, the dynamics of the vocal tract articulators are translated in a waveform that displays fluctuations in signal amplitude over time. This pattern is referred to as the speech signal's envelope, and its main temporal modulation is typically situated between 2-8 Hz, with an average maximum around 4-5 Hz (Ding et al., 2017; Varnet, Ortiz-Barajas, Erra, Gervain, & Lorenzi, 2017). Critically, this characteristic range is preserved across speakers, languages and speaking conditions (Ding et al., 2017; Poeppel & Assaneo, 2020). Speech thus appears to be temporally structured, a feature that the brain might capitalize on to further process relevant information (Rathcke, to appear, this volume). Multiple temporal modulations are crucial for comprehension, including those within the 1-7 Hz range related to phrases, words, and syllables (Elliott & Theunissen, 2009; Meyer, 2018). Temporal modulations above 12 Hz are linked to specific phonetic features and segmental information (Zhang, Zou, & Ding, to appear, this volume; Christiansen, Greenberg, Christiansen, & Bygning, 2009; Drullman, Festen, & Plomp, 1994; Rosen, 1992; Shannon et al., 1995). Speech signals lacking the naturally occurring envelope temporal modulations are less intelligible (Chi, Gao, Guyton, Ru, & Shamma, 1999; Chi, Ru, & Shamma, 2005; Elhilali, Chi, & Shamma, 2003; Elliott & Theunissen, 2009). Moreover, removing the main temporal fluctuations (2–9 Hz) within spoken stimuli by artificially filtering the signal, results in degraded intelligibility for listeners. And artificially restoring these temporal modulations —by the addition of brief noise bursts that act as temporal cues at exactly where the 'acoustic edges' of the original stimuli were— leads to a drastic increase in intelligibility (Doelling, Arnal, Ghitza, & Poeppel, 2014; Ghitza, 2012).

## 3. NEURAL TRACKING OF THE SPEECH ACOUSTIC DYNAMICS

At the neural level, the auditory cortex effectively represents the speech envelope (Nourski et al., 2009; Shamma, 2001). Theta band (4–8 Hz) neural activity consistently aligns with the speech envelope, which closely approximates the syllabic time scale (Rimmele and Keitel, to appear, this volume; Giraud & Poeppel, 2012; Luo & Poeppel, 2007). However, theta activity primarily encodes acoustic rather than linguistic features (Etard & Reichenbach, 2019). Although crucial for intelligibility, the speech envelope indeed only indirectly reflects syllabic rate, which is rather landmarked by acoustic onset edges (Schmidt et al., 2021; Zhang, Zou, & Ding, 2023; Oganian & Chang, 2019). Neural tracking of the speech envelope is hence a necessary, but not sufficient, condition for comprehension (Ahissar et al., 2001; Brodbeck & Simon, 2020; Anne Kösem, Dai, McQueen, & Hagoort, 2023). Additionally, while natural speech's temporal modulation rate is around 5 Hz, neural processes can adapt to acoustic rates up to 15 Hz, beyond which comprehension is hindered (Giroud et al., 2023).

This channel capacity associated with the acoustic modulation rate has a strong impact on speech comprehension, and is independent from syllabic or any other linguistic features. Taken together, the above results converge to reveal the central role of the temporal envelope in speech processing.

## 4. THE SYLLABIC TIME SCALE

The hierarchical structure of language implies the existence of different linguistic units which are combined in different ways to create an infinite number of meanings. While currently there is no consensus on the nature of the fundamental unit of speech recognition, it is generally accepted that features described in phonetics are at work during language perception. Two pre-lexical levels of description have been subject to intense neurophysiological investigation due to their relevance for speech perception: phoneme-sized units (either of a phonetic or a phonemic nature) and syllable-sized units (Giraud & Poeppel, 2012; Mesgarani, Cheung, Johnson, & Chang, 2014; Poeppel & Assaneo, 2020).

Syllables last between 150 and 300 ms, with an average around 200 ms (Ghitza & Greenberg, 2009; Greenberg, 2001; Rosen, 1992). This corresponds to a rate of 2.5–8 syllables per second in natural settings (Coupé et al., 2019; Kendall, 2013; Pellegrino et al., 2011; Zhang et al., 2023). The syllable is an essential unit of all languages, with regard to acquisition, pathologies, language errors and psycholinguistic processing (Rathcke, to appear, this volume; Dolata, Davis, & Macneilage, 2008).

Accordingly, the syllabic timescale is the strongest linguistic determinant of speech comprehension (Giroud et al., 2023). Previous research using speeded speech has provided evidence that beyond 15 syllables per second speech becomes unintelligible (Dupoux & Green, 1997; Foulke & Sticht, 1969; Ghitza, 2014; Giroud et al., 2023; Nourski et al., 2009). 15 Hz would hence be the channel capacity associated with syllabic processing (Giroud et al., 2023). Further use of time-compressed spoken materials showed that compressing natural speech three times or more impairs comprehension but that this effect is strongly alleviated by the insertion of periods of silence between time-compressed speech segments (Ghitza & Greenberg, 2009). In particular, restoring the "syllabicity" of the spoken stimuli (its original temporal structure in terms of the syllable rate), seems to be the optimal way to partially restore comprehension of highly compressed speech. Overall, this suggests that the online tracking of individual syllables is a strong prerequisite for speech comprehension.

## 5. THE PHONEMIC TIME SCALE

While the syllabic time scale and its neural underpinning has been investigated in depth, the contribution and neural substrate of the phonemic time scale to speech comprehension is less clear. Phonemes are the smallest linguistic units of speech sounds and represent a generalization or abstraction over different phonetic realizations. Phonemes are the smallest perceptual unit capable of determining the meaning of a word (e.g., *beer* and *peer* differ only with respect to their initial phonemes). They last typically between 60 and 150 ms in natural speech, with the majority being around 50-80 ms (Ghitza & Greenberg, 2009; Rosen, 1992). This corresponds to a rate of approximately 10-15 phonemes per second in natural speech (Studdert-Kennedy, 1986). Phonemes are associated with a processing bottleneck whose channel capacity is of ~35 Hz (Giroud et al., 2023). However, the phonemic rate has only a residual impact on speech comprehension (Giroud et al., 2023), which suggests that the online tracking of individual phonemes is not a prerequisite for speech comprehension. Instead, acoustic phonetic representations of speech are encoded during natural speech perception (Mesgarani et al., 2014; Nourski et al., 2015), with multiple speech sounds being encoded in parallel at any given time, together with their relative order within the speech sequence (Gwilliams, King, Marantz, & Poeppel, 2022).

Phonemes are however a relevant unit of representation for speech processing. During their first months, infants have the ability to discriminate most phonemic contrasts present in multiple language (Gervain, 2015; Mahmoudzadeh et al., 2013; Moon, Lagercrantz, & Kuhl, 2013) and about six months after birth this ability becomes more focused on native phonemes (Kuhl, 2000; Kuhl, Ramírez, Bosseler, Lin, & Imada, 2014). Interestingly, this specialization in processing native phonemes is linked to an increase in synchronization of low-gamma band (~25-50 Hz) neural activity (Ortiz-Mantilla, Hämäläinen, Realpe-Bonilla, & Benasich, 2016; see Menn, Männel, & Meyer, 2023 for a perspective). In adults, neural dynamics in the low-gamma band are also observed during auditory processing (Lakatos et al., 2005; Lehongre, Ramus, Villiermet, Schwartz, & Giraud, 2011; Morillon, Liégeois-Chauvel, Arnal, Bénar, & Giraud, 2012; Morillon et al., 2010) and notably track the amplitude envelope of speech (Di Liberto, O'Sullivan, & Lalor, 2015; Fontolan, Morillon, Liegeois-Chauvel, & Giraud, 2014; Gross et al., 2013; Lehongre et al., 2011; Lizarazu, Lallier, & Molinaro, 2019). Whether this phenomenon reflects phonemic-categorical processing or lower-level acoustic or phonetic processing remains unclear. Work by Marchesotti and colleagues (Marchesotti et al., 2020) provides evidence of the crucial role played by low-gamma band neural dynamics in processing phonemic information during speech perception. In their study, they recorded electroencephalography (EEG) data from dyslexic participants and found that activity at 30 Hz was lower than that of neurotypical adults. They then used transcranial alternating current stimulation (tACS) to temporarily restore low-gamma neural dynamics in dyslexic adults. Interestingly, this intervention led to improved phonological processing and reading performance, but only when the stimulation was targeted at 30 Hz (vs. 60 Hz) and in the group of participants with dyslexia. These findings support a connection between low-gamma neural oscillations and phonological processing.

## 6. Higher-Level Linguistic Time Scales

Phonemes and syllables are combined to form larger units such as words, phrases and sentences. The length, variability and rhythmicity of these higher-level linguistic structures have been investigated (Breen, 2018; Clifton, Carlson, & Frazier, 2006). These (post-)lexical timescales, however, are of the same order of magnitude as prosodic dynamics, making their specific investigation difficult (but see Section D, to appear, this volume). In particular, in spoken languages, prosodic information (intonation, pauses) naturally fluctuates around 0.5-3 Hz, which encompasses phrasal and word-level timescales (Auer, Couper-Kuhlen, & Muller, 1999; Ghitza, 2017; Inbar, Grossman, & Landau, 2020; Stehwien & Meyer, 2021). Such speech dynamics are tracked by neural dynamics in the same range, which corresponds to the delta frequency band (Bonhage, Meyer, Gruber, Friederici, & Mueller, 2017; Boucher, Gilbert, & Jemel, 2019; Bourguignon et al., 2013; Buiatti, Peña, & Dehaene-Lambertz, 2009; Gross et al., 2013; Meyer, Henry, Gaston, Schmuck, & Friederici, 2017; Molinaro, Lizarazu, Lallier, Bourguignon, & Carreiras, 2016; Park, Ince, Schyns, Thut, & Gross, 2015). The distinctive role of these delta rate dynamics in the temporal cortex for prosodic tracking and high-level linguistic processes has been documented (Bourguignon et al., 2013; Ding, Melloni, Zhang, Tian, & Poeppel, 2016; Keitel, Gross, & Kayser, 2018; Anne Kösem & van Wassenhove, 2017; Lamekina & Meyer, 2022; Lu, Jin, Ding, & Tian, 2023; Molinaro & Lizarazu, 2018; Rimmele, Sun, Michalareas, Ghitza, & Poeppel, 2023;

Vander Ghinst et al., 2016), but their respective channel capacity remains to be explored. Of note, this phrasal tracking occurs even in the absence of distinct acoustic modulations at the phrasal rate (Ding et al., 2016; Kaufeld et al., 2020; Keitel et al., 2018). However, diverging evidence led to the proposal to dissociate delta neural activity driven by acoustically-driven segmentation following prosodic phrases from activity that indexes knowledge-based segmentation of semantic/syntactic phrases (Lu et al., 2023; Meyer, 2018). Currently, the role of delta neural dynamics in speech processing is still vigorously debated (Boucher et al., 2019; Giraud, 2020; Inbar et al., 2020; Kazanina & Tavano, 2023; Lo et al.,2023).

Strikingly, during speech perception, spontaneous finger tapping at the perceived (prosodic) rhythm of speech occurs within the delta range (i.e. at ~2.5 Hz, c.f. (Lidji, Palmer, Peretz, & Morningstar, 2011). A similar effect is visible during music perception, with spontaneous movements occurring at the perceived beat, around 0.5-4 Hz (Merchant, Grahn, Trainor, Rohrmeier, & Fitch, 2015; Morillon, Arnal, Schroeder, & Keitel, 2019; Rajendran, Teki, & Schnupp, 2018). These findings point toward a preference of attentional and motor systems for the slow (~0.5-3 Hz) temporal dynamics of auditory streams. Accordingly, during speech processing delta oscillations are not only visible in temporal areas, but also in the motor cortex (Giordano et al., 2017). And delta motor cortical dynamics uniquely contribute to both the modulation of auditory processing and comprehension: On the one hand, the tracking of acoustic dynamics by the (left) auditory cortex is principally modulated by motor areas, through delta (and to a lesser extent theta) oscillatory activity (Keitel, Ince, Gross, & Kayser, 2017; Park et al., 2015). On the other hand, in motor areas, both delta-tracking of the phrasal acoustic rate and delta-beta coupling predicts speech comprehension (Keitel et al., 2018).

## 7. FROM SPEECH RATE TO INFORMATION RATE

Past works have characterized the properties of language in terms of informational content exchange and transmission using large cross-linguistic corpora and the information theory framework (Shannon, 1948). In this context, 'information' does not refer to message meaning but to its unpredictability or unexpectedness (Coupé et al., 2019; Oh, Coupé, Marsico, & Pellegrino, 2015; Pellegrino et al., 2011). Pellegrino and colleagues (Coupé et al., 2019; Pellegrino et al., 2011) investigated how effectively different languages convey information, positing that languages globally share similarities due to human cognitive architecture. They calculated the syllabic information rate, quantifying the average information per syllable transmitted per second. Their studies revealed that many languages exhibit comparable channel capacity associated with syllabic processing, as evidenced by similar syllabic information rates. However, different strategies were visible across languages, captured by the visible trade-off between information density and speech rate across languages (Coupé et al., 2019; Pellegrino et al., 2011). In other words, in some languages, such as Japanese, speakers tend to pronounce a lot of syllables per second (~8), with each syllable being mildly informative, while in other languages, such as Thai, speaker pronounce less syllable per second (~5), but each syllable is more informative. Overall, the amount of syllabic information transmitted per second is comparable across languages. This suggests that languages have adapted to fit the temporal constraints imposed by the processing bottleneck of syllabic information. Extending this research to online speech comprehension, Giroud and colleagues showed that both phonemic and syllabic information rates impose a processing bottleneck which significantly limits speech comprehension. However, these informational features were found to have a smaller impact on comprehension than higher-level lexical and supra-lexical information (Giroud et al., 2023).

At the lexical level, listeners take advantage of one of the most striking properties of language: the fact that all words do not have the same probability to be uttered. Indeed, words obey a Zipfian distribution (Zipf, 1935), which characterizes the frequency at which they occur in natural language (as computed from a corpora of millions of words). 'The' is the most common English word, while 'persiflage' occurs rarely. The word frequency highly correlates with the mean duration needed to recognize a word (Howes & Solomon, 1951). Accordingly, compressed sentences with a higher density of unexpected words are more difficult to understand (Giroud et al., 2023) and the rate at which this lexical information (or 'static lexical surprise', derived from word frequency) occurs during speeded speech perception is a major determinant of speech comprehension, independently to the previously described lower-level linguistic features.

## 8. Contribution of Contextual Information

The channel capacity of contextual (acoustic or lexical) information —the maximum amount of contextual information that listeners can process per unit of time— can also be determined. Contextual information in speech refers to the additional information from surrounding sounds, words or sentences that can be used to guide perception and enhance comprehension. For instance, listeners can make use of the acoustic context (e.g., specific acoustic cues such as fundamental frequency or voice-onset time) to adaptively and predictively process speech in specific situations (Idemaru & Holt, 2011; Zhang, Wu, & Holt, 2021; Lamekina & Meyer, 2022). Furthermore, not only the nature but also the timing of events is highly relevant for comprehension. For instance, contextual speech rate has been shown to affect the detection of subsequent words (Dilley & Pitt, 2010; Kösem et al., 2018), word segmentation boundaries (Reinisch, Jesse, & McQueen, 2011), and perceived constituent durations (Bosker, 2017).

Listeners capitalize also on contextual lexical information to process speech. Sentences with less expected endings (containing a surprising last word, as in 'the little red riding camembert') result in a larger negative deflection of the electroencephalographic (EEG) signal 400 ms after the onset on the closing word: the classical N400 component (Kutas & Hillyard, 1984). Thanks to the ever growing availability of large language models —these are deep neural networks trained on language material in an unsupervised way— researchers now have access to models that capture the statistical properties of the language data they are trained on, at different levels of the linguistic hierarchy. This enables a finer characterization of the contextual information contained in large corpora that reflect what listeners should expect during everyday communication situations.

By comparing the neural activity patterns evoked by different linguistic units to the probabilities assigned to those units by large language models, researchers can gain insights into the nature of the mental representations of linguistic features in the brain (Brodbeck, Hong, & Simon, 2018; Caucheteux & Gramfort, 2021; Donhauser & Baillet, 2020; Frank, Otten, Galli, & Vigliocco, 2015; Goldstein et al., 2022; Heilbron, Armeni, Schoffelen, Hagoort, & de Lange, 2022; Schrimpf et al., 2021). Combining a deep neural network (GPT-2) to estimate contextual predictions with neural recordings from participants that were listening to audiobooks, Heilbron and colleagues found that brain responses are continuously modulated by linguistic predictions. They observed an impact of contextual predictions at the level of meaning, grammar, words, and speech sounds, and found that high-level predictions can inform low-level ones (Heilbron et al., 2022). Contextual predictions at the word level (i.e., contextual lexical information) extracted from GPT-2 also linearly maps onto the brain responses to speech (Caucheteux, Gramfort, & King, 2023). Overall, these results link predictive coding and language processing frameworks into a coherent picture (but see (Antonello & Huth, 2022) for a conflicting view).

At the behavioral level, contextual lexical surprise (i.e., the unexpectedness of a word given the sentence context which was extracted from a large language model) strongly impacts comprehension of compressed sentences (Giroud et al., 2023). Critically, in natural speech and at normal speed, the intrinsic statistics associated with contextual lexical information are already close to its channel capacity. This suggests that contextual lexical surprise is an important constraint regarding the rate at which natural speech unfolds.

## 9. Exploring Linguistic Levels and their Associated Neural Mechanisms within the Channel Capacity Framework

The work presented here highlights multiple linguistic units and their respective timescales which are relevant for speech processing. They encompass different levels of organizational description and complexity, from acoustic to supra-lexical. More precisely, we characterized the speech acoustic time scales; the syllabic and phonemic time scales; higher-level linguistic time scales such as words, phrases and sentences; lexical information rate and contextual information (Figure 1). While it appears that they each contribute to speech comprehension individually, it is likely that they also interact in a complex way in natural speech conditions to define a global channel capacity associated with our comprehension system (Giroud et al., 2023). For instance, high-level contextual lexical information drives lexical access during continuous speech perception (Gwilliams et al., 2022), lexical information modulates in turn phonological processing via the maintenance of sub-phonemic details in auditory cortex over hundreds of milliseconds (Gwilliams, Linzen, Poeppel, & Marantz, 2018), and when prior context constrains lexical processing, sub-lexical representations are inhibited as they are no longer as important for further processing (Martin, 2016, 2020).

A way forward in deepening our understanding of the neural processes at play during speech comprehension is to develop formal descriptions and measurements of computational units and test their relevance experimentally. In this view, a path worth exploring is the pursuit of the development of more and more precise models of speech and language processing using artificial neural networks (Arana, Pesnot Lerousseau, & Hagoort, 2023; Dunbar et al., 2022). Studying the learned representations of these models can provide insights into meaningful representations for speech comprehension without relying on linguistic concepts (Dunbar et al., 2022). Previous research has demonstrated that the retrieved model representations have similar spectro-temporal parameters as those measured directly in the human auditory cortex (Riad, Karadayi, Bachoud-Lévi, & Dupoux, 2021). In silico models offer several other advantages, including the ability to train them under specific conditions and stimuli and observe their resulting behaviors (Kanwisher, Khosla, & Dobs, 2023). These models can also be used to make testable predictions and hypotheses about speech processing in the brain, thus guiding the development of new theories of language processing and acquisition. For instance, Caucheteux and colleagues determined that large language model's representations about upcoming words can be used to predict brain activity more accurately than representations from preceding words. Moreover, enhancing this algorithm with predictions that span multiple words improves this brain mapping, and these predictions are organized hierarchically, with frontoparietal cortices predicting higher-level, longer-range and more contextual representations than temporal cortices (Caucheteux et al., 2023). Such a result, specifically the exact depth of representations, would have been difficult to predict with such precision solely through theoretical models or experimental paradigms. Another interesting property of these models is that they can be used to select highly specific stimuli (word, sentences) that result in specific model behavior (e.g., strong response of the layers or suppressed response). These stimuli can then be presented to participants while their brain activity is recorded to observe the neural response of the network supporting language processing (Tuckute et al., 2024).

We believe that having a framework that combines the modeling approach with standard experimental methodologies can lead to new insights into the mechanisms underlying comprehension. To that end, we propose that a sensible extension to previous natural language processing (NLP) studies —that have primarily focused on examining comprehension during listening to spoken utterances that fall within the range of typical everyday communication scenarios— would be to explore speech comprehension through the lens of the channel capacity framework. This approach involves pushing the comprehension system to its limits by presenting listeners with speech signals that are difficult to understand, in order to identify the specific acoustic and linguistic features that are crucial for comprehension, their associated channel capacity, and how such processing bottlenecks are implemented in neural dynamics. Large language models offer an unprecedented level of resolution in describing the features of language and speech signals at multiple scales. This opens up new opportunities for researchers to gain a more detailed understanding of which specific features and time scales are crucial for comprehension.

## 10. CONCLUSION

In conclusion, speech is a highly complex signal structured at various levels of analysis. Because of its multiplexed nature, the necessary computations and neural circuits involved in speech processing are likely to be spatially and temporally highly organized. Throughout this chapter, we have examined the temporal constraints limiting speech comprehension beyond the acoustic level (see also Mandke, to appear, this volume). How these hierarchical bottlenecks are implemented, what determines their channel capacity and how they interact to efficiently process speech is currently unknown. We have also demonstrated the potential of the channel capacity framework in enhancing our comprehension of speech processing in humans. As such, developing a research program aiming at determining the capacity limitations of our cognitive resources for comprehension could be instrumental in developing predictive and remediative strategies for improving comprehension skills. One potential avenue is to tailor speech materials or specific interventions to individual cognitive resources to increase the efficiency of information transmission and encoding, thus reducing miscomprehension.

*Summary*

Speech is a complex signal that contains different levels of information at distinct timescales, from acoustic to supra-lexical. This chapter highlights the importance of multiple linguistic features to understand human comprehension ability. The temporal dynamics of these levels of analysis is discussed, along with how they fit with neural data.

*Implications*

Each linguistic feature can be expressed in a number of units per second and their associated channel capacity can be derived. These channel capacities are temporal constraints for speech comprehension and can shape the multiplexed rhythms that are observed in speech and language.

*Gains*

The approach put forward in this chapter lays the foundation for deeper investigations into how the temporal unfolding of multi-level linguistic features impacts speech comprehension. We encourage the use of a normative framework (the concept of channel capacity) to explore the neural mechanisms of speech and language processing.

*Index terms*

Channel capacity, Temporal constraints, Linguistic features, Neural dynamics, Speech processing, Speech rate, Syllabic rate, Phonemic rate, Information rate, Contextual information